\pdfoutput=1
\RequirePackage{ifpdf}
\ifpdf 
\documentclass[pdftex]{sigma}
\else
\documentclass{sigma}
\fi

\numberwithin{equation}{section}

\begin{document}

\newcommand{\arXivNumber}{1311.2240}

\allowdisplaybreaks

\renewcommand{\PaperNumber}{066}

\FirstPageHeading

\ShortArticleName{Non-Point Invertible Transformations and Integrability of Partial Dif\/ference Equations}

\ArticleName{Non-Point Invertible Transformations
\\
and Integrability of Partial Dif\/ference Equations}

\Author{Sergey Ya.~STARTSEV}
\AuthorNameForHeading{S.Ya.~Startsev}
\Address{Ufa Institute of Mathematics, Russian Academy of Sciences,\\
112 Chernyshevsky Str., Ufa, 450077, Russia}
\Email{\href{mailto:startsev@anrb.ru}{startsev@anrb.ru}}

\ArticleDates{Received November 10, 2013, in f\/inal form June 11, 2014; Published online June 17, 2014}

\Abstract{This article is devoted to the partial dif\/ference quad-graph equations that can be represented in the form
$\varphi (u(i+1,j),u(i+1,j+1))=\psi (u(i,j),u(i,j+1))$, where the map $(w,z) \rightarrow (\varphi(w,z),\psi(w,z))$ is
injective.
The transformation $v(i,j)=\varphi (u(i,j),u(i,j+1))$ relates any of such equations to a~quad-graph equation.
It is proved that this transformation maps Darboux integrable equations of the above form into Darboux integrable
equations again and decreases the orders of the transformed integrals by one in the~$j$-direction.
As an application of this fact, the Darboux integrable equations possessing integrals of the second order in
the~$j$-direction are described under an additional assumption.
The transformation also maps symmetries of the original equations into symmetries of the transformed equations (i.e.\
preserves the integrability in the sense of the symmetry approach) and acts as a~dif\/ference substitution for symmetries
of a~special form.
The latter fact allows us to derive necessary conditions of Darboux integrability for the equations def\/ined in the f\/irst
sentence of the abstract.}


\Classification{39A14; 37K05; 37K10; 37K35}

\section{Introduction and basic def\/initions}

\label{sec1}
Let~$u$ be a~function of integers~$i$ and~$j$.
By $u_{p,q}$ denote the shifted value $u(i+p,j+q)$ of this function (in particular, $u:=u_{0,0}=u(i,j)$).
Let $F(u,u_{1,0},u_{0,1})$ be a~single-valued function.
Integrable (in various senses) equations of the form
\begin{gather}
\label{uij}
u_{1,1}=F(u,u_{1,0},u_{0,1})
\end{gather}
are actively studied in recent years (see, e.g.,
\cite{ABS,LS,LY,Mikh,R2001} and references within).
In the present article we mostly consider so-called {\it Darboux integrable} equations, i.e.~equations~\eqref{uij} for
which there exist functions~$I$ and~$J$ such that the relations
\begin{gather}
\label{iib}
I(u_{0,1},u_{1,1},\dots,u_{m,1})=I(u,u_{1,0},\dots,u_{m,0}),
\\
\label{ijb}
J(u_{1,0},u_{1,1},\dots,u_{1,n})=J(u,u_{0,1},\dots,u_{0,n})
\end{gather}
hold true for any solution~$u$ of the equation.
(In other words, relationships~\eqref{iib},~\eqref{ijb} mean that~$I$ and~$J$ remain unchanged after the shifts in~$j$
and~$i$, respectively.) The functions~$I$ and~$J$ are respectively called an~$i$-integral of order~$m$ and
a~$j$-integral of order~$n$.
The discrete wave equation $u_{1,1}=u_{1,0}+u_{0,1}-u$ gives us the simplest example of a~Darboux integrable equation.
Here $I=u_{1,0}-u$ and $J=u_{0,1}-u$.
It should be noted that this work (in contrast to, for instance,~\cite{GY}) deals only with the autonomous integrals,
i.e.~with the integrals that do not depend explicitly on the discrete variables~$i$ and~$j$.

The equations~\eqref{uij} can be considered as dif\/ference analogues of the partial dif\/ferential equations
\begin{gather}
u_{xy}=F(u,u_x,u_y).
\label{hyp}
\end{gather}
The concept of the Darboux integrability was initially introduced for partial dif\/ferential equations back in the 19th century.
(The more recent term~$C$-integrability, which was of\/fered in~\cite{Cal}, in some sense generalizes this concept.)
Searching for Darboux integrable equations of the form~\eqref{hyp} was started in classical works such as~\cite{Guq},
and the most recent and complete classif\/ication result was obtained in~\cite{ZhSok} more than a~century later.
At present, similar classif\/ication results are absent for Darboux integrable equations~\eqref{uij}, only separate
examples (see, for instance,~\cite{GY,HZS2,Sit}) and the description~\cite{Stn} of a~special case with $n=1$
in~\eqref{ijb} are known.
Therefore, it seems reasonable to consider a~subclass of Darboux integrable equations~\eqref{uij} as an intermediate goal.

To def\/ine this subclass, in the present paper we assume that the right-hand side of~\eqref{uij} satisf\/ies the inequalities
\begin{gather}
\label{hipc}
\frac{\partial F}{\partial u} \ne 0,
\qquad
\frac{\partial F}{\partial u_{1,0}} \ne 0,
\qquad
\frac{\partial F}{\partial u_{0,1}} \ne 0
\end{gather}
and a~relationship of the form
\begin{gather}
\label{qua}
\varphi\left(u_{1,0},F(u,u_{1,0},u_{0,1})\right)=\psi(u,u_{0,1}),
\end{gather}
where $\varphi(w,z)$ and $\psi(w,z)$ are functionally independent.
It is easy to see that the left-hand side of~\eqref{qua} is the shift of $\varphi(u,u_{0,1})$ in~$i$ by virtue
of~\eqref{uij}.
If we denote the operator of this shift by~$T_{\rm i}$, then~\eqref{qua} reads as
\begin{gather}
\label{invt}
T_{\rm i}(\varphi (u, u_{0,1})) = \psi(u, u_{0,1}),
\qquad
\frac{\partial \varphi}{\partial u} \frac{\partial \psi}{\partial u_{0,1}} - \frac{\partial \varphi}{\partial u_{0,1}}
\frac{\partial \psi}{\partial u} \ne 0.
\end{gather}
Thus, we consider formal dif\/ference analogues of the quasilinear partial dif\/ferential equations $u_{xy}=a(u,u_y) u_x + b(u,u_y)$
(because any of these dif\/ferential equations can be represented as $D_{x}(\varphi(u,u_y))=\psi(u,u_y)$ in terms of the total derivative $D_x$).
For shorter formulations, below we refer to equation~\eqref{invt} for designating an equation of the form~\eqref{uij}
that satisf\/ies~\eqref{qua}.

In a~part of the article we use the stronger assumptions that the map $(w,z)\!\rightarrow\!
(\varphi(w,z),\psi(w,z))$ is injective, the integral~$J$ has second order and $\varphi(u,u_{0,1})$ is
uniquely expressed in terms of~$J$ and $\varphi(u_{0,1},u_{0,2})$.
(Since we prove below that~$j$-integrals are expressed in terms of $\varphi(u,u_{0,1})$ and its shifts in~$j$, the last
assumption is not so restrictive as it seems; moreover, it can be omitted if a~certain general statement will be
proved~-- see the last paragraph of this section for more details.)
It should be noted that one of the most known
integrable equations on the quad-graph, the discrete Liouville equation
\begin{gather}
\label{dle}
u_{1,1}=\frac{(u_{1,0}-1)(u_{0,1}-1)}{u}
\end{gather}
from~\cite{Hirota}, satisf\/ies all of the above assumptions.
Indeed, the work~\cite{AdS} demonstrates that
\begin{gather*}
I=\left(\frac{u_{2,0}}{u_{1,0}-1} +1 \right) \left(\frac{u-1}{u_{1,0}} + 1 \right),
\qquad
J=\left(\frac{u_{0,2}}{u_{0,1}-1} +1 \right) \left(\frac{u-1}{u_{0,1}} + 1 \right)
\end{gather*}
for~\eqref{dle}, the corresponding functions $\varphi(w,z)=z/(w-1)$, $\psi(w,z)=(z-1)/w$ def\/ine the injective map and
$\varphi(u,u_{0,1})=(\varphi(u_{0,1},u_{0,2})+1)/(J - \varphi(u_{0,1},u_{0,2}) - 1)$.
Thus, the present article is devoted to `nearest relatives' of the discrete Liouville equation~\eqref{dle}.
More precisely, we focus on interactions of a~non-point invertible transformation~\cite{Sit} with Darboux integrability
and, to a~lesser extent, higher symmetries of the quad-graph equations.

This transformation is def\/ined in the following way.
We can rewrite~\eqref{invt} in the form of the system
\begin{gather}
\label{absd}
v=\varphi (u,u_{0,1}),
\qquad
v_{1,0}:=T_{\rm i}(v)=\psi (u,u_{0,1}),
\end{gather}
express~$u$ and $u_{0,1}$ in terms of~$v$, $v_{1,0}$ from~\eqref{absd} and obtain
\begin{gather}
\label{pqsd}
u=\Omega(v,v_{1,0}),
\qquad
u_{0,1}=\Upsilon(v,v_{1,0}),
\end{gather}
where~$\Omega$ and~$\Upsilon$ are functionally independent.
According to~\eqref{hipc}, the functions~$\varphi$ and~$\psi$ are assumed here to be essentially depending on both their
arguments, and~\eqref{pqsd} therefore implies that~$\Omega$ and~$\Upsilon$ essentially depend on both their arguments
too.
The system~\eqref{pqsd} is equivalent to the equation
\begin{gather}
\label{pqd}
\Omega(v_{0,1},v_{1,1})=\Upsilon(v,v_{1,0}).
\end{gather}
Generally speaking, the above procedure is well-def\/ined only locally and the right-hand sides of~\eqref{pqsd} may be
dif\/ferent for dif\/ferent pairs $(u,u_{0,1})$.
We avoid this if no more than one pair $(u,u_{0,1})$ satisf\/ies the system~\eqref{absd} for any given~$v$, $v_{1,0}$.
Under this assumption, the equation~\eqref{pqd} is well-def\/ined and the transformation $v=\varphi (u,u_{0,1})$ maps all
solutions of~\eqref{invt} into solutions of~\eqref{pqd}.
Because the set of solutions to~\eqref{pqd} may be wider than the image of solutions to~\eqref{invt} under the
transformation $v=\varphi (u,u_{0,1})$, we can not guarantee that the inverse transformation $u=\Omega(v,v_{1,0})$ maps
any solution of~\eqref{pqd} into a~solution of~\eqref{invt} (some pairs $(v,v_{1,0})$ may generate another equations of
the form~\eqref{invt} when we perform the above procedure in the inverse order).
But we can restore all equations related to~\eqref{pqd} by using the inverse procedure.
The described transformation is the direct analogue of that was of\/fered in~\cite{Yam90} for dif\/ferential-dif\/ference
equations.
Some applications of the transformation~\eqref{absd}--\eqref{pqd} can also be found in~\cite{Sit,Yam94}.

It is almost obvious that this transformation preserves Darboux integrability.
But a~formal proof of this fact is still needed to demonstrate, for example, that no~$j$-integral of~\eqref{invt}
becomes constant after substituting $u=\Omega(v,v_{1,0})$ into it.
Such a~proof is given in Section~\ref{s2}.
More precisely, we prove that equation~\eqref{pqd} has a~$j$-integral of order $n-1$ and an~$i$-integral of order $m+1$
if equation~\eqref{invt} possesses~$j$- and~$i$-integrals of orders~$n$ and~$m$, respectively.
This reduces the classif\/ication problem for Darboux integrable equations~\eqref{invt} admitting~$n$-th
order~$j$-integrals to the classif\/ication of equations~\eqref{pqd} possessing~$j$-integrals of order $n-1$.
As an example of such kind, in Section~\ref{s2} we completely describe the Darboux integrable equations~\eqref{invt}
admitting a~second-order~$j$-integral such that $\varphi(u,u_{0,1})$ is uniquely expressed in terms of this integral and
$\varphi(u_{0,1},u_{0,2})$.

In Section~\ref{s3} we study the interaction between the transformation~\eqref{absd}--\eqref{pqd} and symmetries of the
quad-graph equations.
Analogically to the case of semi-discrete equations~\cite{Yam90}, the transformation $v=\varphi (u,u_{0,1})$ def\/ines
a~dif\/ference substitution for special symmetries of~\eqref{invt} and, under additional assumptions, maps any higher
symmetry of~\eqref{invt} into a~higher symmetry of~\eqref{pqd} (i.e.~the transformation preserves integrability in the
sense of the symmetry test\footnote{This test is described, for example, in~\cite{GY,LY}.}).
The former fact allows us to obtain necessary conditions of Darboux integrability for the equations of the
form~\eqref{invt} if~$u$ is uniquely expressed in terms of $\varphi (u,u_{0,1})$ and $u_{0,1}$.

Now, let us introduce notation and more formal def\/initions.

Due to the conditions~\eqref{hipc}, we can express any argument of the right-hand side~$F$ of~\eqref{uij} in terms of
the others and rewrite this equation, after appropriate shifts in~$i$ and~$j$, in any of the following forms
\begin{gather}
\label{umm}
u_{-1,-1}=\overline{F}(u,u_{-1,0},u_{0,-1}),
\\
\label{upm}
u_{1,-1}=\hat{F}(u,u_{1,0},u_{0,-1}),
\\
\label{ump}
u_{-1,1}=\tilde{F}(u,u_{-1,0},u_{0,1}).
\end{gather}
These formulas (and their consequences derived by shifts in~$i$ and~$j$) allow us to express any `mixed shift'
$u_{p,q}$, $p q \ne 0$, in terms of $u_{k,0}$, $u_{0,l}$ at least locally (i.e.~in an enough small neighborhood of any
arbitrary selected solution of~\eqref{uij}, which is considered in a~f\/inite number of the points $(i, j)$).
Therefore, we can formulate our reasonings only in terms of an arbitrary solution~$u$ to~\eqref{uij} and its `canonical
shifts' $u_{k,0}$, $u_{0,l}$, $k,l \in \mathbb{Z}$.
These `canonical shifts' are called {\it dynamical variables} and can be considered as functionally independent.
(The mixed shift elimination procedure and the dynamical variables are described in more details, for example,
in~\cite{Mikh}.) We use the notation $g[u]$ to designate that the function~$g$ depends on a~f\/inite number of the
dynamical variables.
All functions are assumed to be analytical in this paper, and our considerations are local.

In general, the right-hand sides of~\eqref{umm}--\eqref{ump} are not uniquely def\/ined (may vary with~$i$,~$j$ and with
a~solution in a~neighborhood of which we consider these functions).
This does not matter if we use~\eqref{umm}--\eqref{ump} to only estimate what dynamical variables do local expressions
for $u_{p,q}$ depend on.
But this is important in certain cases, and our statements therefore contain the single-valuedness assumptions for
$\hat{F}$ when it is needed (to avoid dif\/f\/iculties discussed, for example, in~\cite{NeA}).

Let $T_{\rm i}$ and $T_{\rm j}$ denote the operators of the forward shifts in~$i$ and~$j$ by virtue of the
equation~\eqref{uij}.
For any function~$f$, they satisfy the rules
\begin{gather*}
T_{\rm i}(f(a,b,c,\dots))=f(T_{\rm i}(a),T_{\rm i}(b),T_{\rm i}(c),\dots),
\\
T_{\rm j} (f(a,b,c,\dots))=f(T_{\rm j} (a),T_{\rm j} (b),T_{\rm j} (c),\dots).
\end{gather*}
In addition, $T_{\rm i}$ and $T_{\rm j}$ map $u_{p,q}$ into $u_{p+1,q}$ and $u_{p,q+1}$, respectively, and then replace
any mixed shift of~$u$ with its expression in terms of the dynamical variables.
For example,
\begin{gather}
\label{u1n}
T_{\rm i}(u_{0,n})=T_{\rm j}^{n-1}(F),
\qquad
T_{\rm i}(u_{0,-n})=T_{\rm j}^{1-n}(\hat{F}),
\qquad
n \in \mathbb{N}.
\end{gather}
Here a~shift operator with a~superscript~$k$ designates the~$k$-fold application of this operator (e.g.
$T_{\rm j}^3:= T_{\rm j} \circ T_{\rm j} \circ T_{\rm j}$, $T_{\rm i}^{-2}:= T_{\rm i}^{-1} \circ T_{\rm i}^{-1}$ and
any operator with the zero superscript is the identity mapping).
The inverse (backward) shift operators $T_{\rm i}^{-1}$ and $T_{\rm j}^{-1}$ are def\/ined in a~similar way.

\begin{definition}
\label{def1}
Let functions $I[u]$ and $J[u]$ satisfy the relations $T_{\rm j}(I)=I$ and $T_{\rm i}(J)=J$ for an equation of the
form~\eqref{uij}, and let each of the functions essentially depend on at least one of the dynamical variables.
Then $I[u]$ and $J[u]$ are respectively called an~$i$-integral and a~$j$-integral of the equation~\eqref{uij}, and this
equation is called Darboux integrable.
\end{definition}
It is easy to prove that the~$i$- and~$j$-integrals have the form $I(u_{k,0},u_{k+1,0},\dots,u_{k+m,0})$ and
$J(u_{0,l},u_{0,l+1},\dots,u_{0,l+n})$, respectively (see, for example,~\cite{Stn}).
The numbers~$m$ and~$n$ are called {\it order} of the corresponding integral if $I_{u_{k,0}} I_{u_{k+m,0}} \ne 0$ and
$J_{u_{0,l}} J_{u_{0,l+n}} \ne 0$.
We can set $k=l=0$ without loss of generality because $T^{-k}_{\rm i}$ and $T^{-l}_{\rm j}$ respectively map any~$i$-
and~$j$-integrals into~$i$- and~$j$-integrals again.
Thus, equations~\eqref{umm}--\eqref{ump} are in fact not needed for the above def\/inition.

\begin{definition}
An equation $u_{t} = f[u]$ is called a~symmetry of equation~\eqref{uij} if the relation $L(f)=0$ holds true, where
\begin{gather*}
L = T_{\rm i} T_{\rm j} - \frac{\partial F}{\partial u_{1,0}} T_{\rm i} - \frac{\partial F}{\partial u_{0,1}} T_{\rm j}
- \frac{\partial F}{\partial u}.
\end{gather*}
\end{definition}
According to~\cite{AdS,Stn}, if equation~\eqref{uij} is Darboux integrable and uniquely solvable for $u_{1,0}$ (i.e.~the
right-hand side $\hat{F}$ of~\eqref{upm} is uniquely def\/ined), and $J[u]$ denotes its~$j$-integral, then there exists an
operator
\begin{gather}
\label{rop}
R=\sum\limits_{q=0}^r \lambda_q (u_{0,\varrho},u_{0,\varrho +1},\dots,u_{0,s}) T^q_{\rm j},
\qquad
\lambda_r \ne 0,
\end{gather}
such that
\begin{gather}
\label{ed}
u_t=R\big(\eta(T_{\rm j}^{p}(J), T_{\rm j}^{p+1}(J), T_{\rm j}^{p+2}(J), \dots)\big)
\end{gather}
is a~symmetry of this equation for any integer~$p$ and any function~$\eta$ depending on a~f\/inite number of the
arguments.
Both the present paper and the article~\cite{Stn} (results of which we use below) in fact describe equations that
admit~$j$-integrals and symmetries of the form~\eqref{ed} (the existence of~$i$-integrals is not really used in the most
part of the reasonings).
We therefore make additional assumptions on a~$j$-integral in Proposition~\ref{d2p} and $\hat{F}$ in
Corollaries~\ref{dc1},~\ref{dc2} to guarantee that the transformed and the original equations are uniquely solvable for
$v_{1,0}$ and $u_{1,0}$, respectively.
Proposition~\ref{d2p} and Corollaries~\ref{dc1},~\ref{dc2} will remain valid without these assumptions if the existence
of symmetries~\eqref{ed} for the Darboux integrable equations is proved without employing the single-valuedness of
$\hat{F}$.

\section{Transformation of integrals}
\label{s2}

It is convenient for further reasoning to prove the following proposition f\/irst.
\begin{lemma}
\label{l0}
All functions in the set $\{T_{\rm i}^p(\varphi), T_{\rm j}^q(\varphi) \,|\, p,q \in \mathbb{Z}, q \ne 0 \}$ are
functionally independent for any equation of the form~\eqref{invt}.
\end{lemma}
\begin{proof}
The function~$\psi$ can be rewritten as $\eta_1(\varphi,u)$, where $\eta_1$ must depend on its second argument due to
the functional independence of~$\varphi$ and~$\psi$.
Using the formula $T_{\rm i}(\varphi)=\psi=\eta_1(\varphi,u)$ and induction on~$p$, we obtain that $T_{\rm
i}^p(\varphi)=\eta_p(\varphi,u,u_{1,0},\dots,u_{p-1,0})$ depends on $u_{p-1,0}$ for any $p>1$.
Also,~$\varphi$ can be represented as a~function $\eta_{-1}(\psi,u)$ that depends on its second argument.
Therefore, $T_{\rm i}^{-1}(\varphi)=\eta_{-1}(\varphi,u_{-1,0})$ and $T_{\rm
i}^p(\varphi)=\eta_p(\varphi,u_{-1,0},\dots,u_{p,0})$ depends on $u_{p,0}$ for $p<0$.
Thus, the functions $T_{\rm i}^p(\varphi)$ are functionally independent because~$\varphi$ and $T_{\rm i}(\varphi)=\psi$
are functionally independent and any other $T_{\rm i}^p(\varphi)$ depends on the variable that is absent in either all
previous (if $p>1$) or all next (if $p<0$) members of the sequence $T_{\rm i}^s(\varphi)$.

If $q>0$, then the function $T_{\rm j}^q(\varphi)=\varphi(u_{0,q},u_{0,q+1})$ can not be expressed in terms of $T_{\rm
i}^p(\varphi)$, $p \in \mathbb{Z}$, and $T_{\rm j}^s(\varphi)$, $s<q$, because they do not depend on $u_{0,q+1}$.
If $q<0$, then the function $T_{\rm j}^q(\varphi)=\varphi(u_{0,q},u_{0,q+1})$ can not be expressed in terms of $T_{\rm
i}^p(\varphi)$, $p \in \mathbb{Z}$, and $T_{\rm j}^s(\varphi)$, $s>q$, because they do not depend on $u_{0,q}$.
Thus, $\{T_{\rm i}^p(\varphi), T_{\rm j}^q(\varphi) \,|\, p,q \in \mathbb{Z}, q \ne 0 \}$ is a~set of functionally
independent functions.
\end{proof}

\begin{lemma}
\label{intf}
Up to shifts in~$j$, any~$n$-th order~$j$-integral of equation~\eqref{invt} can be represented in the form
$\Phi(\varphi(u,u_{0,1}),\varphi(u_{0,1},u_{0,2}),\dots,\varphi(u_{0,n-1},u_{0,n}))$.
This representation is well-defined on solutions of~\eqref{uij}, and the function~$\Phi$ essentially depends on its
first and last arguments.
\end{lemma}
\begin{proof}
As it is demonstrated in the comments to Def\/inition~\ref{def1}, we can assume without loss of generality that
the~$j$-integral has the form $J(u,u_{0,1},\dots,u_{0,n})$.
Equation~\eqref{qua} implies that the right-hand side of~\eqref{uij} has the form $g(\psi(u,u_{0,1}),u_{1,0})$,
where~$g$ is single-valued.
The backward shift in~$i$ gives us $u_{0,1}=g(\varphi(u,u_{0,1}),u)$.
Using this expression and its consequences derived by shifts in~$j$, we rewrite the~$j$-integral as
$J=\Phi(u,\varphi(u,u_{0,1}),\varphi(u_{0,1},u_{0,2}),\dots,\varphi(u_{0,n-1},u_{0,n}))$.
Since
\begin{gather*}
T_{\rm i} \big(\Phi(u, \varphi,\dots,T_{\rm j}^{n-1}(\varphi))\big) = \Phi\big(u_{1,0}, \psi,\dots,T_{\rm j}^{n-1}(\psi)\big)
\end{gather*}
by virtue of~\eqref{invt}, the relation $T_{\rm i}(\Phi)=\Phi$ can hold true only if~$\Phi$ does not depend on its f\/irst
argument ($\Phi_u=0$).
The function~$\Phi$ also must depends on~$\varphi$ and $T_{\rm j}^{n-1}(\varphi)$ because $J_{u}=0$ and $J_{u_{0,n}}=0$
otherwise.
\end{proof}

\begin{theorem}
\label{t1}
Let the map $(w,z) \rightarrow (\varphi(w,z),\psi(w,z))$ be injective, and the corresponding
equation~\eqref{pqd} be uniquely solvable for $v_{1,1}$.
Then the equation~\eqref{invt} possesses an~$i$-integral of order~$m$ and a~$j$-integral of order~$n$ if and only if the
equation~\eqref{pqd} possesses an~$i$-integral of order $m+1$ and a~$j$-integral of order $n-1$.
If $n=2$ and~\eqref{pqd} is not uniquely solvable for~$v_{1,1}$, then the Darboux integrability of~\eqref{invt} implies
the existence of an $(m+1)$th order~$i$-integral and a~first-order~$j$-integral for an equation
$v_{1,1}=Q(v,v_{1,0},v_{0,1})$ that satisfies the relationship
$\Omega(v_{0,1},Q(v,v_{1,0},v_{0,1}))=\Upsilon(v,v_{1,0})$.
\end{theorem}
\begin{proof}
We can assume without loss of generality that the integrals of~\eqref{invt} have the form $I(u,u_{1,0},\dots,u_{m,0})$
and $J(u,u_{0,1},\dots,u_{0,n})$.
Substituting $\Omega(\varphi,T_{\rm i}(\varphi))$ instead of~$u$ into the~$i$-integral, we obtain
$I(\Omega,\Omega_1,\dots,\Omega_m)$, where $\Omega_k=\Omega (T_{\rm i}^k(\varphi),T_{\rm i}^{k+1}(\varphi))$.
The relationship $T_{\rm j}(I)=I$ takes the form
\begin{gather}
\label{ipq}
I(\Upsilon,\Upsilon_1,\dots,\Upsilon_m)=I(\Omega,\Omega_1,\dots,\Omega_m),
\qquad
\Upsilon_k=\Upsilon\big(T_{\rm i}^k(\varphi),T_{\rm i}^{k+1}(\varphi)\big),
\end{gather}
because $v=\varphi(u,u_{1,0})$ satisf\/ies equation~\eqref{pqd} and hence $T_{\rm j}$ maps $\Omega(\varphi,T_{\rm
i}(\varphi))$ into $\Upsilon(\varphi,T_{\rm i}(\varphi))$.
But~$\varphi$, $T_{\rm i}(\varphi)$, $\dots$, $T_{\rm i}^{m+1}(\varphi)$ are functionally independent by Lemma~\ref{l0},
and~\eqref{ipq} can be valid only if the relation
\begin{gather*}
I(\Upsilon(v,v_{1,0}),\Upsilon(v_{1,0},v_{2,0}),\dots,\Upsilon(v_{m,0},v_{m+1,0}))
\\
\qquad
=I(\Omega(v,v_{1,0}),\Omega(v_{1,0},v_{2,0}),\dots,\Omega(v_{m,0},v_{m+1,0}))
\end{gather*}
holds true identically for arbitrary~$v, v_{1,0}, \dots, v_{m+1,0}$.
And this means that the right-hand side of the last relationship is an~$i$-integral of equation~\eqref{pqd}.

According to Lemma~\ref{intf}, $J=\Phi(\varphi,\dots,T_{\rm j}^{n-1}(\varphi))$.
Let us rewrite~\eqref{pqd} as $v_{1,1}=Q(v,v_{1,0},v_{0,1})$.
Since $T_{\rm i}(\Phi)=T_{\rm i}(J)=J=\Phi$ and equation~\eqref{pqd} holds true for $v=\varphi(u,u_{0,1})$, the function
$\Phi(v,v_{0,1},\dots,v_{0,n-1})$ satisf\/ies the def\/ining relation
\begin{gather}
\label{defe}
\Phi\big(v_{1,0},Q,\dots,T_{\rm j}^{n-2}(Q)\big)=\Phi(v,v_{0,1},\dots,v_{0,n-1})
\end{gather}
for~$j$-integral of~\eqref{pqd} when $v_{1,0}=T_{\rm i}(\varphi)$ and $v_{0,k}=T_{\rm j}^k(\varphi)$,
$k=\overline{0,n-1}$ (other variables are absent in the def\/ining relation).
But $T_{\rm i}(\varphi)=\psi$ and $T_{\rm j}^k(\varphi)$ are functionally independent and, therefore,~\eqref{defe} is
valid only if it holds true identically for arbitrary $v_{1,0}$ and $v_{0,k}$, $k=\overline{0,n-1}$.
Thus, $\Phi(v,v_{0,1},\dots,v_{0,n-1})$ is a~$j$-integral of equation~\eqref{pqd}.

 Conversely, let $I(\Omega(v,v_{1,0}),\Omega(v_{1,0},v_{2,0}),\dots,\Omega(v_{m,0},v_{m+1,0}))$ and
$\Phi(v,v_{0,1},\dots,v_{0,n-1})$ be inte\-grals of~\eqref{pqd}.
Since $v=\varphi(u,u_{1,0})$ satisf\/ies~\eqref{pqd} for any solution~$u$ of the equation~\eqref{invt} and
$u=\Omega(\varphi,T_{\rm i}(\varphi))$ identically holds true, the relationships
\begin{gather*}
T_{\rm i}\big(\Phi(\varphi,T_{\rm j}(\varphi),\dots,T_{\rm j}^{n-1}(\varphi))\big)
=\Phi\big(\varphi,T_{\rm j}(\varphi)\dots,T_{\rm j}^{n-1}(\varphi)\big),
\\
T_{\rm j}(I(u,u_{1,0},\dots,u_{m,0}))=I(u,u_{1,0},\dots,u_{m,0})
\end{gather*}
follow from the def\/ining relation for integrals of~\eqref{pqd}.
\end{proof}

In particular, this theorem implies that any Darboux integrable equation~\eqref{invt} admitting
a~second-order $j$-integral
can be derived from a~Darboux integrable equation possessing a~f\/irst-order~$j$-integral.
But the equations of the latter kind were described (under an additional assumption) in the recent work~\cite{Stn} and
we only need to select the equations of the form~\eqref{pqd} among them.

\begin{lemma}
\label{ll3}
Let an equation
\begin{gather}
\label{fpr}
\tilde{v}_{1,1}=Q(\tilde{v},\tilde{v}_{1,0},\tilde{v}_{0,1}),
\qquad
\frac{\partial Q}{\partial \tilde{v}} \frac{\partial Q}{\partial \tilde{v}_{1,0}} \frac{\partial Q}{\partial
\tilde{v}_{0,1}} \ne 0
\end{gather}
be Darboux integrable, satisfy a~relationships of the form
$\tilde{\Omega}(\tilde{v}_{0,1},Q(\tilde{v},\tilde{v}_{1,0},\tilde{v}_{0,1}))=\tilde{\Upsilon}(\tilde{v},\tilde{v}_{1,0})$
and possess a~first-order~$j$-integral $\Phi(\tilde{v},\tilde{v}_{0,1})$ such that the equation
$\Phi(\tilde{v},\tilde{v}_{0,1})=\tilde{w}$ is uniquely solvable for $\tilde{v}$.
Then a~point transformation $\tilde{v}=\zeta(v)$ relates~\eqref{fpr} to an equation of the form
\begin{gather}
\label{strt}
\frac{\delta D - v_{1,1} (A v_{0,1} + C - \delta B)}{v_{1,1}-v_{0,1}} = \frac {(v+B)(\delta v_{1,0} +C)-AD}{v_{1,0}-v},
\end{gather}
where the constants~$A$,~$B$,~$C$,~$D$ and~$\delta$ satisfy the inequalities $|\delta A| + |C - \delta B| + |D| \ne 0$,
$|\delta|+|A|+|C| \ne 0$.
\end{lemma}
\begin{proof}
The def\/ining relation $\Phi(\tilde{v}_{1,0}, Q) =\Phi(\tilde{v},\tilde{v}_{0,1})$ for the~$j$-integral~$\Phi$ is
uniquely solvable for the f\/irst argument of the left-hand side by the theorem assumptions.
Hence, the equation~\eqref{fpr} is uniquely solvable for $\tilde{v}_{1,0}$.
According to~\cite{Stn}, an equation of the form~\eqref{fpr} is Darboux integrable, uniquely solvable for
$\tilde{v}_{1,0}$ and admits an autonomous f\/irst-order~$j$-integral only if a~point change of variables
$\tilde{v}=\zeta(v)$ relates this equation to an equation of the form
\begin{gather}
\label{veq}
v_{1,1}=\alpha (\phi) + \frac{\beta(\phi)}{\gamma(\phi)-v_{1,0}},
\qquad
\beta \ne 0,
\end{gather}
where~$\phi$ is the function of~$v$ and $v_{0,1}$ that satisf\/ies the relationship
\begin{gather}
\label{veq2}
v_{0,1}=\alpha (\phi) + \frac{\beta(\phi)}{\gamma(\phi)-v}.
\end{gather}
It is obvious that the last relationship can hold true only if $|\alpha'|+|\beta'|+|\gamma'| \ne 0$.
The function~$\phi$ is a~$j$-integral, and
\begin{gather}
\label{gin}
I[v] = \frac{v_{3,0}-v_{1,0}}{v_{3,0}-v_{2,0}} \cdot \frac{v_{2,0}-v}{v_{1,0}-v}
\end{gather}
is an~$i$-integral of~\eqref{veq}.
It should be noted that some equations of the form~\eqref{veq}, \eqref{veq2} admit~$i$-integrals of order less than 3,
but all such equations possess the integral~\eqref{gin} too (see~\cite{Stn} for more details).

An equation $v_{1,1}=F(v,v_{1,0},v_{0,1})$ can be written in the form~\eqref{pqd} only if the condition
$(F_v/F_{v_{1,0}})_{v_{0,1}} =0$ holds true.
Substituting the right-hand side of~\eqref{veq} into this condition, we obtain
\begin{gather}
\label{cond}
\frac{\partial}{\partial v_{0,1}} \left(\frac{\alpha' (\gamma - v_{1,0})^2 + \beta' (\gamma - v_{1,0}) - \beta
\gamma'}{\beta} \frac{\partial \phi}{\partial v} \right) = 0.
\end{gather}
On the other hand, dif\/ferentiation of~\eqref{veq2} with respect to~$v$ gives rise to
\begin{gather*}
\beta= \big(\beta \gamma' - \alpha' (\gamma - v)^2 - \beta' (\gamma - v)\big) \frac{\partial \phi}{\partial v}.
\end{gather*}
The condition $\beta \ne 0$ implies that the both factors in the right-hand side do not equal zero.
Therefore,~\eqref{cond} takes the form
\begin{gather*}
\frac{\partial}{\partial v_{0,1}} \left(\frac{\alpha' (\gamma - v_{1,0})^2 + \beta' (\gamma - v_{1,0}) - \beta
\gamma'}{\alpha' (\gamma - v)^2 + \beta' (\gamma - v) - \beta \gamma'} \right) = 0.
\end{gather*}
Performing the dif\/ferentiation in the left-hand side of the last relationship, we obtain
\begin{gather*}
\big[(\alpha' \gamma'' -\alpha'' \gamma ') \big(v_{1,0}^2 - 2 \gamma v_{1,0}\big) + \big(\gamma' \beta'' - \gamma'' \beta' + 2 \alpha'
(\gamma')^2\big) v_{1,0}\big] \beta
\\
\qquad
{}+ \big(\alpha'' \beta' - \alpha' \beta'' - 2 (\alpha')^2 \gamma'\big)\big(v_{1,0}^2(\gamma - v) + v_{1,0} \big(v^2 -\gamma^2\big)\big) + \dots=0,
\end{gather*}
where the dots signify terms without $v_{1,0}$.
The left-hand side is a~polynomial in~$v$ and $v_{1,0}$.
The coef\/f\/icients of it depend on~$\phi$ only and must be equal to zero because $\phi(v,v_{0,1})$,~$v$ and $v_{1,0}$ are
functionally independent.
Thus, we have
\begin{gather}
\label{abg}
\alpha' \gamma'' -\alpha'' \gamma' = \gamma' \beta'' - \gamma'' \beta' + 2 \alpha' (\gamma')^2 = \alpha' \beta'' -
\alpha'' \beta' + 2 (\alpha')^2 \gamma' = 0.
\end{gather}

If~$\phi$ satisf\/ies~\eqref{veq2}, then any function of~$\phi$ also satisf\/ies~\eqref{veq2} (with
another~$\alpha$,~$\beta$ and~$\gamma$).
This is why we can assume without loss of generality that $\alpha=\delta \phi$ if $\alpha' \ne 0$.
Under this assumption,~\eqref{abg} gives rise to $\gamma=A \phi - B$ and $\beta=D - C \phi -\delta \phi (A \phi - B)$.
It is easy to check that an appropriate change $v \rightarrow v+\lambda$ in~\eqref{veq2} reduces the two other cases
$\alpha'=0$, $\gamma = \phi$ and $\alpha' = \gamma' =0$, $\beta = \phi$ to the same formulas for~$\alpha$,~$\beta$
and~$\gamma$ with $\delta=0$, $A=1$ and $\delta=A=D=0$, $C=-1$, respectively.
Substituting these formulas into~\eqref{veq2} and solving it for~$\phi$, we obtain
\begin{gather}
\label{phi}
\phi = \frac{v_{0,1}(v + B)+D}{A v_{0,1} + \delta v+C}.
\end{gather}
The corresponding equation~\eqref{veq} is
\begin{gather*}
v_{1,1}=\frac{((v+B)(\delta v_{1,0} +C) - AD)v_{0,1} + \delta D (v_{1,0}- v)}{A(v_{1,0}-v) v_{0,1} + (v_{1,0}+B)(\delta
v +C) - AD}
\end{gather*}
and can be rewritten as~\eqref{strt}.
\end{proof}

If we perform a~point transformation $v=\eta(\tilde{v})$ in~\eqref{pqd}, then the corresponding system
\begin{gather*}
u=\Omega(\eta(\tilde{v}),\eta(\tilde{v}_{1,0})),
\qquad
u_{0,1}=\Upsilon(\eta(\tilde{v}),\eta(\tilde{v}_{1,0}))
\end{gather*}
has the solution $\eta(\tilde{v})=\varphi(u,u_{0,1})$, $\eta(\tilde{v}_{1,0})=\psi(u,u_{0,1})$ that generates the
unchanged equation~\eqref{invt}.
Thus, we can restore the original equation~\eqref{invt} after a~point transformation in~\eqref{pqd}.
But the equation $\tilde{\Omega}(\tilde{v}_{0,1},\tilde{v}_{1,1})=\tilde{\Upsilon}(\tilde{v},\tilde{v}_{1,0})$ in the
formulation of Lemma~\ref{ll3} may also dif\/fer from~\eqref{pqd} via a~transformation $\tilde{\Omega}=\xi(\Omega)$,
$\tilde{\Upsilon}=\xi(\Upsilon)$, that corresponds to the point change $\tilde{u}=\xi(u)$ in~\eqref{invt}.

Theorem~\ref{t1}, the above lemmas and the reasonings of the previous paragraph mean that the transformation
from~\eqref{pqd} to~\eqref{invt} for equation~\eqref{strt} gives us all, up to point transformations, Darboux integrable
equations~\eqref{invt} admitting second-order~$j$-integrals of the form
$\Phi(\varphi(u,u_{0,1}),\varphi(u_{0,1},u_{0,2}))$, where the equation $\Phi(v,v_{0,1})=w$ is uniquely solvable
for~$v$.
Let us perform this transformation.

The system~\eqref{pqsd} for~\eqref{strt} takes the form
\begin{gather}
\label{u10}
u=\frac{\delta D - v_{1,0} (A v + C - \delta B)}{v_{1,0}-v},
\\
\label{u11}
u_{0,1} = \frac {(v+B)(\delta v_{1,0} +C) - AD}{v_{1,0}-v}.
\end{gather}
Equation~\eqref{u10} can be rewritten in the form
\begin{gather}
\label{uint}
v_{1,0}=\frac{u v + \delta D}{u + A v + C - \delta B}.
\end{gather}
Replacing $v_{1,0}$ with the right-hand side of~\eqref{uint}, we transform~\eqref{u11} into the algebraic equation $P_2
v^2 + P_1 v + P_0 =0$, where
\begin{gather}
\label{p2}
P_2 = \delta u + A u_{0,1} + A C,
\\
\label{p1}
P_1 = (C + \delta B) u + (C - \delta B) u_{0,1} + (B C - A D - \delta D) (A - \delta) + C^2,
\\
\label{p0}
P_0 = (B C - A D) u - \delta D u_{0,1} + B C (C - \delta B) + D \big(\delta A B - A C + \delta^2 B\big).
\end{gather}
Solving this equation for~$v$ and substituting a~solution $\theta (u,u_{0,1})$ into~\eqref{uint}, we obtain the
following classif\/ication result.

\begin{proposition}
\label{d2p}
Let the map $(w,z) \rightarrow (\varphi(w,z),\psi(w,z))$ be injective, and let the equation~\eqref{invt}
possess a~second-order autonomous~$j$-integral~$J$ such that $\varphi(u,u_{0,1})$ is uniquely expressed in terms of~$J$
and $\varphi(u_{0,1},u_{0,2})$.
Then the equation~\eqref{invt} is Darboux integrable if and only if a~point change of variables $\rho(u) \rightarrow u$
reduces this equation to the form
\begin{gather}
\label{ggen}
T_{\rm i}(\theta (u,u_{0,1}))=\frac{u \theta (u,u_{0,1}) + \delta D}{A \theta(u,u_{0,1}) + u + C - \delta B},
\end{gather}
where~$\theta$ is a~solution of the equation $P_2 \theta^2 + P_1 \theta + P_0=0$ with the coefficients defined by
formulas~\eqref{p2}--\eqref{p0}, and the constants~$A$,~$B$,~$C$,~$D$,~$\delta$ satisfy the conditions $|\delta|+|A|+|C|
\ne 0$, $|\delta A| + |C - \delta B| + |D| \ne 0$.
\end{proposition}
Theorem~\ref{t1} guaranties that~\eqref{ggen} is Darboux integrable, and the proof of Theorem~\ref{t1} gives us the way
to construct integrals of this equation.
Substituting~$\theta$ instead of~$v$ into~\eqref{phi}, we obtain a~$j$-integral
\begin{gather*}
J[u]=\phi(\theta,T_{\rm j}(\theta))=\frac{T_{\rm j}(\theta)(\theta + B)+D}{A T_{\rm j}(\theta) + \delta \theta+C}.
\end{gather*}
Equation~\eqref{uint} (and its shifts in~$i$) allows us to represent $v_{1,0}$, $v_{2,0}$, $v_{3,0}$ as functions
of~$v$,~$u$, $u_{1,0}$, $u_{2,0}$.
Replacing $v_{k,0}$ with these functions in~\eqref{gin}, we derive the formula
\begin{gather*}
I[u]= \frac{(u_{2,0}+u_{1,0} + C - \delta B) (u_{1,0}+u + C - \delta B)}{u_{1,0}^2 + u_{1,0} (C - \delta B) - \delta
A D}
\end{gather*}
for an~$i$-integral of the equation~\eqref{ggen}.

As it is shown in Section~\ref{sec1}, the discrete Liouville equation~\eqref{dle} satisf\/ies all assumptions of
Proposition~\ref{d2p}.
Therefore, it lies in the equation family~\eqref{ggen} as a~particular case $C = -1$, $B \ne 0$, $A=\delta=0$.
Another special case $C=D=0$, $B=-1$, $A \ne -1$, $\delta =1$ gives us the equation $u_{1,1} (u+1) = u_{1,0} (u_{0,1}
+1)$, which is contained in a~list of Darboux integrable equations in~\cite{GY}.

\section{Transformation of symmetries}
\label{s3}
Let $h_*$ designate the Frech\'et derivative (linearization operator)
\begin{gather*}
h_* = \sum\limits_{q=-\infty}^{+\infty} \frac{\partial h}{\partial u_{q,0}} T^q_{\rm i} +
\sum\limits_{\genfrac{}{}{0pt}{}{q=-\infty}{q \ne 0}}^{+\infty} \frac{\partial h}{\partial u_{0,q}} T^q_{\rm j}
\end{gather*}
of the function $h[u]$, and let $\partial_f$ denote the dif\/ferentiation with respect to~$t$ by virtue of the equation
$u_t=f[u]$.
The formula ${\partial_f (h[u]) = h_* (f)}$ def\/ines $\partial_f$ on the functions of the dynamical variables.

Let us consider a~dif\/ferential-dif\/ference equation of the form
\begin{gather}
\label{eg}
u_t=g(u_{0,n},u_{0,n+1},\dots,u_{0,k}).
\end{gather}

\begin{definition}
\label{drp}
If a~function $\phi (u_{0,l},u_{0,l+1},\dots,u_{0,m})$, $m > l$, satisf\/ies the inequality $\phi_{u_{0,l}} \phi_{u_{0,m}}
\ne 0$ and a~relationship\footnote{This relationship means that $v=\phi[u]$ maps solutions of~\eqref{eg} into solutions
of the equation $v_t=\hat{g}$.} of the form
\begin{gather}
\label{dpd}
\partial_g (\phi) = \hat{g} \big(T^n_{\rm j}(\phi), T^{n+1}_{\rm j}(\phi), \dots, T^k_{\rm j}(\phi)\big),
\end{gather}
then we say that equation~\eqref{eg} admits the dif\/ference substitution
\begin{gather}
\label{rp}
v=\phi (u_{0,l},u_{0,l+1},\dots,u_{0,m})
\end{gather}
into the equation $v_t=\hat{g}(v_{0,n},v_{0,n+1},\dots,v_{0,k})$.

We call~\eqref{rp} a~Miura-type substitution if there exist operators~\eqref{rop} and
\begin{gather*}
\hat{R}=\sum\limits_{q=l}^{r+m} \hat{\lambda}_q (v_{0,\hat{\varrho}},v_{0,\hat{\varrho}+1},\dots,v_{0,\hat{s}}) T^q_{\rm
j}
\end{gather*}
such that the equation $u_t=R(\xi (T^{p}_{\rm j} (\phi), T^{p+1}_{\rm j} (\phi), \dots))$ admits the
substitution~\eqref{rp} into the equation $v_t=\hat{R} (\xi (v_{0,p},v_{0,p+1},\dots))$ for any integer~$p$ and any
function $\xi$ depending on a~f\/inite number of the arguments.
\end{definition}
Since mixed shifts $u_{q,p}$, $q p \ne 0$, do not appear in the def\/ining relation~\eqref{dpd}, the above def\/inition in
no way uses equation~\eqref{uij}.
However, equations~\eqref{uij} can generate dif\/ference substitutions, for example, in the way that was used
in~\cite{Yam90} for dif\/ferential-dif\/ference analogues of equations~\eqref{invt}.

\begin{theorem}
\label{t2}
If~\eqref{eg} is a~symmetry of~\eqref{invt}, then the equation~\eqref{eg} admits the difference substitutions
$v=\varphi(u,u_{0,1})$ and $v=\psi(u,u_{0,1})$ into an equation $v_t=\hat{g}(v_{0,n},v_{0,n+1},\dots,v_{0,k})$.

If the map $(w,z) \rightarrow (\varphi(w,z),\psi(w,z))$ is injective and the corresponding
equation~\eqref{pqd} is uniquely solvable for $v_{1,0}$, $v_{1,1}$ and $v_{0,1}$, then the following proposition is also
true for any symmetry $u_t=f[u]$ of equation~\eqref{invt} such that $f\ne 0$.
Let $\hat{f}[v]$ designate the function $\varphi_*(f)$ after substituting $u=\Omega(v,v_{1,0})$ into it and excluding
the variables of the form $v_{p,q}$, $p q \ne 0$, by virtue of~\eqref{pqd}.
Then $\hat{f}[v] \ne 0$ and $v_t = \hat{f}[v]$ is a~symmetry of equation~\eqref{pqd}.
If, in addition, $f_{u_{\delta,0}} \ne 0$ for some $\delta>0$ $(\delta<0)$ or $f_{u_{0,\sigma}} \ne 0$ for some
$\sigma>0$ ($\sigma < 0$), then $\hat{f}[v]$ essentially depends on $v_{d,0}$ for some $d \ge \delta$ ($d \le \delta$)
or on $v_{0,b}$ for some $b \ge \sigma$ ($b \le \sigma$), respectively.
\end{theorem}

Although the proof is almost identical to the proof of the analogous proposition for semi-discrete equations
in~\cite{Yam90}, we include it for the sake of completeness.
It should also be noted that for the rest part of the present article we need only the f\/irst sentence of the theorem.
\begin{proof}
The application of the Frech\'et derivative to both sides of~\eqref{qua} gives rise to
\begin{gather*}
T_{\rm i} \left(\frac{\partial \varphi}{\partial u_{0,1}} \right)\! F_{*} + T_{\rm i} \left(\frac{\partial
\varphi}{\partial u} \right) = \psi_{*}
\
\Rightarrow
\
T_{\rm i} \left(\frac{\partial \varphi}{\partial u_{0,1}} \right)\! L = T_{\rm i} \left(\frac{\partial \varphi}{\partial
u_{0,1}} \right)\! (T_{\rm i} T_{\rm j} - F_{*}) = T_{\rm i} \circ \varphi_{*} - \psi_{*},
\end{gather*}
where $\circ$ denotes the composition of operators.
Thus, the def\/ining relation $L(f)=0$ for the symmetry $u_t=f[u]$ takes the form $T_{\rm i} (\varphi_{*} (f))= \psi_{*}(f)$.

Let~\eqref{eg} be a~symmetry of~\eqref{invt}.
The function $\varphi_*(g)$ can be rewritten as
\begin{gather}
\label{dgf}
\varphi_*(g)=\hat{g}\big(u_{0,n},T^n_{\rm j}(\varphi),T^{n+1}_{\rm j}(\varphi),\dots,T^k_{\rm j}(\varphi)\big).
\end{gather}
The substitution of~\eqref{dgf} into the def\/ining relation $\psi_{*} (g) = T_{\rm i} (\varphi_{*} (g))$ gives rise to
\begin{gather}
\label{dgp}
\psi_{*} (g) = \hat{g}\big(T_{\rm i}(u_{0,n}),T^n_{\rm j}(\psi),T^{n+1}_{\rm j}(\psi),\dots,T^k_{\rm j}(\psi)\big).
\end{gather}
But~\eqref{hipc} and~\eqref{u1n} imply that $T_{\rm i}(u_{0,n})$ depends on $u_{1,0}$ whereas $\psi_{*} (g)$ and
$T^n_{\rm j}(\psi), \dots, T^k_{\rm j}(\psi)$ do not.
Therefore, $\hat{g}$ does not depend on its f\/irst argument and~\eqref{dgf}, \eqref{dgp} take the form
\begin{gather*}
\varphi_*(g)=\hat{g}\big(T^n_{\rm j}(\varphi),T^{n+1}_{\rm j}(\varphi),\dots,T^k_{\rm j}(\varphi)\big),
\\
\psi_{*} (g) = \hat{g}\big(T^n_{\rm j}(\psi),T^{n+1}_{\rm j}(\psi),\dots,T^k_{\rm j}(\psi)\big),
\end{gather*}
respectively.

For the symmetries of the general form $u_t=f[u]$, equation~\eqref{pqsd} and the relationship $T_{\rm i} (\varphi_{*}
(f))= \psi_{*} (f)$ imply
\begin{gather}
\label{invs}
\partial_f (\Omega(\varphi,\psi))= \frac{\partial \Omega}{\partial v} \varphi_*(f) + \frac{\partial \Omega}{\partial
v_{1,0}} \psi_*(f) = \frac{\partial \Omega}{\partial v} \varphi_*(f) + \frac{\partial \Omega}{\partial v_{1,0}} T_{\rm
i} (\varphi_*(f))= \partial_f (u)= f,
\\
\partial_f (\Upsilon(\varphi,\psi))= \frac{\partial \Upsilon}{\partial v} \varphi_*(f) + \frac{\partial
\Upsilon}{\partial v_{1,0}} \psi_*(f) = \frac{\partial \Upsilon}{\partial v} \varphi_*(f) + \frac{\partial
\Upsilon}{\partial v_{1,0}} T_{\rm i} (\varphi_*(f))= \partial_f (u_{0,1})= T_{\rm j}(f).
\nonumber
\end{gather}
The last two formulas, together with $\psi=T_{\rm i}(\varphi)$, mean that the function $\breve{f}[u]:=\varphi_*(f)$
satisf\/ies the relation\-ship
\begin{gather}
T_{\rm j} \left(\frac{\partial \Omega(\varphi,T_{\rm i}(\varphi))}{\partial v} \breve{f} + \frac{\partial
\Omega(\varphi,T_{\rm i}(\varphi))}{\partial v_{1,0}} T_{\rm i} (\breve{f}) \right)
= T_{\rm j}(f)\nonumber
\\
\qquad
=\frac{\partial\Upsilon(\varphi,T_{\rm i}(\varphi))}{\partial v} \breve{f} + \frac{\partial \Upsilon(\varphi,T_{\rm
i}(\varphi))}{\partial v_{1,0}} T_{\rm i} (\breve{f}).\label{defsym}
\end{gather}
Using the identities $u=\Omega(\varphi,T_{\rm i}(\varphi))$, $\Omega(T_{\rm j}(\varphi),T_{\rm i} T_{\rm j}
(\varphi))=\Upsilon(\varphi,T_{\rm i}(\varphi))$ and their consequences derived by shifts in~$i$ and~$j$, we can
represent $\breve{f}[u]$ as a~function $\hat{f}[\varphi]$ of only $T_{\rm i}^p(\varphi)$, $T_{\rm j}^q(\varphi)$, $p,q
\in \mathbb{Z}$, $q \ne 0$, and exclude all mixed shifts $T_{\rm i}^s T_{\rm j}^r(\varphi)$, $sr \ne 0$,
from~\eqref{defsym}.
As a~result of this, in~\eqref{defsym} the operators $T_{\rm i}$ and $T_{\rm j}$ act on $v_{p,0}=T_{\rm i}^p(\varphi)$,
$v_{0,q}=T_{\rm j}^q(\varphi)$ by virtue of~\eqref{pqd}, i.e.~\eqref{defsym} becomes the def\/ining relation for
symmetries of~\eqref{pqd}.
But the functions $T_{\rm i}^p(\varphi)$, $T_{\rm j}^q(\varphi)$ are functionally independent by Lemma~\ref{l0}, and
$\hat{f}[v]$ therefore satisf\/ies this def\/ining relation not only if $v_{p,0}=T_{\rm i}^p(\varphi)$, $v_{0,q}=T_{\rm
j}^q(\varphi)$ but for arbitrary values of $v_{p,0}$, $v_{0,q}$.

If $\varphi_*(f)=0$, then~\eqref{invs} implies $f=0$.
This is why $\hat{f}[v] \ne 0$ if $f \ne 0$.
The same logic proves the dependence of $\hat{f}[v]$ on $v_{d,0}$ if we take the relationship $T_{\rm i} T_{\rm
j}^q(\varphi)= \psi(u_{0,q},u_{0,q+1})$ and the formulas for $T_{\rm i}^p(\varphi)$, $T_{\rm j}^q(\varphi)$ from the
proof of Lemma~\ref{l0} into account.
The dependence of $\hat{f}[v]$ on $v_{0,b}$ follows directly form the relation $\varphi_{u_{0,1}} T_{\rm j} (f) +
\varphi_u f=\hat{f}[v]\big|_{v=\varphi}$ and the fact that~$f$ can not depend on $u_{0,\sigma}$ for positive
or negative~$\sigma$ if $\hat{f}[v]$ does not depend on $v_{0,b}$ for all $b \ge \sigma$ or for all $b \le \sigma$,
respectively.
\end{proof}

\begin{corollary}
\label{dc1}
If an equation of the form~\eqref{invt} is uniquely solvable for $u_{1,0}$ and Darboux integrable, then
$v=\varphi(u,u_{0,1})$ and $w=\psi(u,u_{0,1})$ are Miura-type substitutions.
\end{corollary}
Recall that in the present paper we understand~\eqref{invt} as an equation of the form~\eqref{uij} satisfying the
relationship~\eqref{qua}.
According to this, in Corollaries~\ref{dc1},~\ref{dc2} we assume that the equation~\eqref{uij} is uniquely solvable for
$u_{1,0}$ (i.e.~the right-hand side of the corresponding equation~\eqref{upm} is well-def\/ined).
We can replace this assumption with the more strong assumption that the formula $v=\varphi(u,u_{0,1})$ is uniquely
solvable for~$u$.
The last assumption guarantees that any equation~\eqref{uij} satisfying~\eqref{qua} is uniquely solvable for $u_{1,0}$.
\begin{proof}
It is easy to see that $\hat{R}$ in Def\/inition~\ref{drp} coincides with the operator $\phi_* \circ R =
\sum\limits_{q=l}^{r+m} \breve{\lambda}_q[u] T_{\rm j}^q$ and $v=\phi[u]$ is a~Miura-type substitution if all
$\breve{\lambda}_q[u]$ can be expressed in terms of $T_{\rm j}^b (\phi)$ only.

According to~\cite{AdS,Stn}, if equation~\eqref{uij} is Darboux integrable and uniquely solvable for $u_{1,0}$, then it
possesses symmetries of the form~\eqref{rop}, \eqref{ed}.
Theorem~\ref{t2} implies that the symmetries~\eqref{ed} admit the substitutions $v=\varphi(u,u_{0,1})$ and
$w=\psi(u,u_{0,1})$, i.e.~$\varphi_*(R(\eta))$ and $\psi_* (R(\eta))$ can be respectively expressed in terms of
$v_{0,b}= T_{\rm j}^b (\varphi)$ and $w_{0,b}:=T_{\rm j}^b (\psi)$, $b \in \mathbb{Z}$.
But in the proof of Lemma~\ref{intf} we demonstrate that any~$j$-integral $J[u]$ of~\eqref{invt} can be represented both
as a~function of $T_{\rm j}^b (\varphi)$ and as a~function of $T_{\rm j}^b (\psi)$.
Hence, the same is true for any function~$\eta$ of the~$j$-integrals.
Due to the arbitrariness of~$\eta$ in~\eqref{ed}, this implies that the coef\/f\/icients of the operators $\varphi_* \circ
R$ and $\psi_* \circ R$ can be expressed in terms of $T_{\rm j}^b (\varphi)$ and $T_{\rm j}^b (\psi)$, respectively.
Thus, the equation $u_t=R(\xi)$ admits the substitutions $v=\varphi(u,u_{0,1})$ and $w=\psi(u,u_{0,1})$ for any function
$\xi$ depending on arguments of the forms $T_{\rm j}^b (\varphi)$ and $T_{\rm j}^b (\psi)$, respectively.
\end{proof}

It is proved in~\cite{foi} that $v=\phi(u,u_{0,1})$ is a~Miura-type substitution only if~$\phi$ satisf\/ies the
relationship
\begin{gather*}
\zeta(u_{0,1})=\alpha (\phi) + \frac{\beta(\phi)}{\gamma(\phi)-\zeta(u)},
\end{gather*}
for some functions~$\alpha$,~$\beta$,~$\gamma$ and~$\zeta$ such that $\beta \zeta' \ne 0$.
This implies the following proposition.
\begin{corollary}
\label{dc2}
If an equation of the form~\eqref{invt} is uniquely solvable for $u_{1,0}$ and Darboux integrable, then there exist
functions~$\alpha$,~$\beta$,~$\gamma$,~$\zeta$ and $\hat{\alpha}$, $\hat{\beta}$, $\hat{\gamma}$, $\hat{\zeta}$ such
that
\begin{gather*}
\zeta(u_{0,1})=\alpha (\varphi) + \frac{\beta(\varphi)}{\gamma(\varphi)-\zeta(u)},
\qquad
\beta \zeta' \ne 0,
\\
\hat{\zeta}(u_{0,1})=\hat{\alpha}(\psi) + \frac{\hat{\beta}(\psi)}{\hat{\gamma}(\psi)-\hat{\zeta}(u)},
\qquad
\hat{\beta} \hat{\zeta}' \ne 0.
\end{gather*}
\end{corollary}
Applying $T_{\rm i}$ and $T_{\rm i}^{-1}$ to the both sides of the former and the latter relationships, respectively, we
obtain that any Darboux integrable equation~\eqref{invt} can be represented in the form
\begin{gather*}
\zeta(u_{1,1})=\alpha (\psi) + \frac{\beta(\psi)}{\gamma(\psi)-\zeta(u_{1,0})}
\end{gather*}
as well as in the form
\begin{gather*}
\hat{\zeta}(u_{-1,1})=\hat{\alpha}(\varphi) + \frac{\hat{\beta}(\varphi)}{\hat{\gamma}(\varphi)-\hat{\zeta}(u_{-1,0})}.
\end{gather*}
We can assume either $\zeta(u)=u$ or $\hat{\zeta}(u)=u$ without loss of generality because we can perform either the
point change of variables $\zeta(u)\rightarrow u$ or the point change $\hat{\zeta}(u) \rightarrow u$.

\subsection*{Acknowledgments}

The author thanks the referees for useful suggestions.
This work is partially supported by the Russian Foundation for Basic Research (grant number 13-01-00070-a).

\pdfbookmark[1]{References}{ref}
\LastPageEnding

\end{document}